# Toward Understanding the Use of Centralized Exchanges for Decentralized Cryptocurrency

*Zhixuan Zhou[1], Bohui Shen[2]*

[1] *University of Illinois at Urbana-Champaign*

*zz78@illinois.edu*

[2] *BNU-HKBU United International College*

*r130233082@mail.uic.edu.hk*

## ABSTRACT

Cryptocurrency has been extensively studied as a decentralized financial technology built on blockchain. However, there is a lack of understanding of user experience with cryptocurrency exchanges, the main means for novice users to interact with cryptocurrency. We conduct a qualitative study to provide a panoramic view of user experience and security perception of exchanges. All 15 Chinese participants mainly use centralized exchanges (CEX) instead of decentralized exchanges (DEX) to trade decentralized cryptocurrency, which is paradoxical. A closer examination reveals that CEXes provide better usability and charge lower transaction fee than DEXes. Country-specific security perceptions are observed. Though DEXes provide better anonymity and privacy protection, and are free of governmental regulation, these are not necessary features for many participants. Based on the findings, we propose design implications to make cryptocurrency trading more decentralized.

**Keywords**: Blockchain, Cryptocurrency Exchange, User Experience, Security

## INTRODUCTION

Cryptocurrency built on blockchain, such as Bitcoin and Ethereum, has become increasingly popular. Such features as anonymity of users and regulation-free enable them to disrupt the traditional financial industry. People trade cryptocurrency via both decentralized exchanges (DEXes) and centralized exchanges (CEXes). DEXes such as Uniswap are built on the Ethereum blockchain and are completely free of centralized regulation. On the other hand, CEXes such as Binance and Coinbase run on centralized servers, and more resemble traditional stock exchanges.

Toward understanding user experience and security perception of cryptocurrency exchanges, a key infrastructure supporting cryptocurrency trading yet receiving limited attention from the research community (Jang et al. 2021), we interviewed 15 exchange users in China. It is worth noticing that we recruited exchange users in general, yet all of them mainly relied on CEXes. This echoes with CoinGecko's ranking of cryptocurrency exchanges by trust score, a metric of popularity and security (Jin, 2020), where top-ranked exchanges are all CEXes, such as Binance, OKX, KuCoin, Gate.io and Crypto.com (top 5 as of April $4^{th}$, 2022). The highest-ranked DEXes are Maiar (80th), Cyber DEX (89th), and Uniswap (90th). In general, CEXes dominate the decentralized cryptocurrency market, which seems paradoxical, and may compromise the decentralization nature of cryptocurrency.

Throughout the interviews, we find that users prefer CEXes due to their good overall experience, learnability and low transaction fee. High gas/transaction fee are main concerns of our participants for DEXes. They are not willing to "pay for decentralization" and indicate that they would only turn to DEXes after the fee is lowered. Despite good user experience of CEXes, privacy violations are found common in C2C transactions and KYC (know your customer) process in the initiation of CEX use. CEXes are also susceptible to regulation by the Chinese government. Country-specific mentalities toward (de)centralization and privacy are an interesting phenomenon. For example, many of these participants from China regard sensitive information leakage during KYC as less severe an issue, since they have been used to it, and for them, decentralization is not a necessary pursuit.

Contributions of this study are thus 2-fold. Firstly, we obtain a deep understanding of user experience and security perceptions of cryptocurrency exchanges, especially CEXes. Secondly, we inform future designs of DEXes to make them more usable for users, toward creating a fully decentralized financial world.

# RELATED WORK

## User Experience/Usability of Blockchain and Cryptocurrency

Blockchain has been applied to various fields, from insurance (Gatteschi et al. 2018) to non-fungible tokens (Sharma et al. 2022). User experience is key to the wide adoption of blockchain, especially for novice users. A recent work (Voskobojnikov et al. 2021) identified and qualitatively analyzed 6,859 reviews regarding user experience of five mobile cryptocurrency wallets. For example, the wallet initialization process was found tedious by some app reviewers, and lack of guidance during the setup process made it challenging to create a wallet. Albayati et al. (2021) combined User Experience Questionnaire (UEQ) with usability to comprehend UX and added trust as a significant construct to understand the impact of user trust in cryptocurrency wallets.

Trading through exchanges is the most common way to get cryptocurrency (Kim et al. 2018). However, user study toward understanding user practices with exchanges is insufficient. Jang et al. (2021) asked twenty participants to perform 6 tasks and evaluate 7 items of usability and showed that the blockchain-based cryptocurrency exchange (KDEX) has worse usability than the centralized exchange (Bithumb). We approach the user experience of exchanges with qualitative interviews of Chinese users, relating it to a specific cultural context.

### Security and Trust in Blockchain and Cryptocurrency

In (Abramova et al. 2021), risk perceptions and security behaviors of different types of crypto-asset users such as cypherpunks, hodlers, and rookies are identified through a survey. Regarding trust in blockchain, in (Sas and Khairuddin, 2017), 20 interviews were conducted in Malaysia to understand users' experience with bitcoin and trust challenges. Blockchain's characteristics supporting users' credibility include honesty ensured by decentralization and public ledger's transparency, reputation supported by large companies' interest in bitcoin, ease of use grounded in ease and quick transactions, limited risk due to transactions' low cost, and the decentralized, unregulated nature of blockchain which limits the risk of institutional power abuse. Gaggioli et al. (2019) pointed out that the adoption of blockchain was not only a technological, but also a psychological challenge, which crucially depended on the possibility of creating a trust management approach that matched the underlying distributed communication system.

Despite the key role of exchanges for people to interact with cryptocurrency, and the large number of attacks targeting exchanges (Marella et al. 2021), little research has been conducted on people's security perceptions and trust evaluations of exchanges. We bridge this research gap with an exploratory interview study.

## METHODOLOGY

We recruited interview participants in the authors' personal contact, as well as on WeChat and Weibo. We eventually had 15 participants for interviews. Demographics of the participants are summarized in Table 1.

Interview questions covered the participants' general understanding of cryptocurrency, exchange(s) they are using, and overall experience/learnability/ /transaction fee/regulation/security/privacy of the exchanges. The interviews were conducted by two native Chinese speakers between March and April, 2022, audio-taped with consent, and transcribed for data analysis. We used an AI-powered transcription tool to generate raw transcripts, and manually corrected mistakes.

We started the analysis while the data were being collected. We used thematic coding (Gibbs, 2007) to interpret the interview transcripts and identify emerging themes, and used a mind mapping tool to organize corresponding quotes into a hierarchy of themes. We did the analysis iteratively, and regularly discussed to reach a consensus. In this paper, we will use individual quotes, which were translated into English, to illustrate our points. All quotes are anonymized to protect privacy of the participants.

Table 1: Basic information of study participants (exchange users).

| ID | Gender | Occupation | Year(s) | Exchange(s) | Platform(s) |
|---|---|---|---|---|---|
| P1 | M | Programmer | <1 | Binance/PancakeSwap | Mobile |
| P2 | F | PhD student | 1 | Binance | Mobile/Desktop |
| P3 | F | Undergraduate student | 2 | OKX | Mobile |
| P4 | M | HR/Master student | 6 | Huobi | Mobile |
| P5 | M | PhD student | 4 | Binance/Huobi | Mobile |
| P6 | M | Investment intern | <1 | OKX/Gate.io | Mobile |
| P7 | M | Gym consultant | 1 | Binance | Mobile |
| P8 | F | Master student | 1 | Binance | Mobile |
| P9 | F | Product manager | 2 | OKX/Binance/Uniswap | Mobile |
| P10 | F | Product manager | 1 | Binance /OKX/Gate.io | Mobile |
| P11 | F | Master student | 1 | OKX | Web |
| P12 | M | Master student | 2 | Binance | Mobile |
| P13 | M | PhD student | 3 | Binance/Gate.io/ BitCoke/FTX | Mobile/Web/Desktop |
| P14 | M | Master student | 7 | Binance | Web |
| P15 | F | Researcher | 1 | Binance | Mobile |

## FINDINGS

In this section, we will first briefly report on participants' general experience with cryptocurrency, as a context to understand their exchange usage. Then we'll elaborate on their experiences and perceptions of CEXes.

### The Use of Decentralized Cryptocurrency

Our participants have interacted with cryptocurrency from less than one year (P1, P6) to 7 years (P14). The average self-evaluated expertise of them was 3.54 on a 1-10 scale (1 indicates novice and 10 indicates savvy), leaning toward the novice end.

Generally, they had pleasurable experience with cryptocurrency, and believed in its value. Most of our participants used cryptocurrency for investment purposes, given its high volatility in price. For example, P3 treated cryptocurrency as "a very risky stock." Several participants (P4, P9, P11) also traded cryptocurrency out of curiosity.

**The Use of Centralized Exchanges**

As can be seen from Table 1, most participants (13/15) only used CEXes. P1 and P9 mostly relied on CEXes, and only used DEXes occasionally. A direct reason for the popularity of CEXes was their advertising and brand building efforts to attract new users. As P1 noticed, Huobi would "pay influencers ("大V") in the cryptocurrency field on Weibo and let them recommend Huobi to newcomers."

To understand why novice users inclined to use CEXes, we explored their user experience and security perception of CEXes in terms of overall experience, learnability, transaction fee, regulation, security, and privacy. We also introduced the concept of DEXes to them, and probed their preference between DEXes and CEXes.

**Overall Experience.** The participants had good overall experience with the CEXes they used. For example: (P3, OKX) "I think it's very professional, I mean, the whole platform seems very professional. The function and the user interface are good. The experience of using it is very happy and smooth."

**Learnability.** User interface (UI) of CEXes resembled that of stock exchanges, thus users could easily get used to them, as pointed out by P9. P10 shared a similar opinion and added that the UI design of Binance well accommodated the needs and habits of Chinese users. CEXes were easy to learn even for people without prior experience in trading. P2 gave credit to Binance's rich educational resources in forms of text, video and interactive tutorials: "It's easy to learn. It will show you how to buy USDT or bitcoin or other digital currency… If you click on some buttons, you'll be directed into some information pages." OKX similarly provided educational materials to "teach users how to change CNY into digital money, and how to buy/sell." (P3)

Many of our participants were not frequent traders, thus they did not need to learn more complicated functions such as leveraged trading. P8 noted that Binance users could shift between Lite and Professional versions in the app, and the Lite version only included basic functions like buy/sell, well suiting novice users' needs.

**Transaction Fee.** Generally, transaction fees on CEXes were much lower than DEXes. As told by some participants, the transaction fee rate on Uniswap was 0.875%, while that on Binance was only 0.1%. P1 acknowledged that he would only consider DEXes if one day the transaction fee was lowered in DEXes.

**Regulation.** CEXes were susceptible to governmental regulations. The participants repeatedly mentioned restrictions on cryptocurrency trading in China, e.g., Internet connection issue, and being denied transactions by banks and payment applications (e.g., Alipay). P6 regarded this as a paradox: "While the crypto community puts a heavy emphasis on decentralization, most strict regulations are put on exchanges." The regulations put great mental burdens on the participants. P8 said, "I'm very concerned that one day the country will stop cryptocurrency trading completely, and I can't withdraw my assets from the exchanges."

Chinese users could not access CEXes without a VPN, which set a barrier for people to download and use them. P2 talked about the difficulties when downloading and registering at the beginning: "When I tried to download the Binance app, I needed to use the VPN. However, when I download it using an American IP address, I could not register with a Chinese mobile phone number. So, it took me quite some time to solve this problem." In addition, every time the CEXes needed upgrading, users had to use a VPN to download them again.

It was also challenging to buy USDT using Chinese Yuan (CNY), or convert USDT to CNY. For example, Alipay would stop P2's transaction, and check if she was trading cryptocurrency: "The customer service center will call me and check if I'm doing something with digital currencies. And I say no. If I'm lucky, I can succeed in my second purchase." P10 avoided such inconvenience by doing fiat-cryptocurrency trading with friends instead of C2C sellers.

Trading cryptocurrency was like a gray zone in China. There were official regulations prohibiting cryptocurrency trading, but they were not enforced so strictly. P5 told the author that many users, especially on Huobi, got their bank accounts frozen, for their cryptocurrency-fiat transactions. In extreme cases, CEXes stopped serving active Chinese users without compensating them after being caught by the police. Unfortunate users like P8 suffered from financial losses.

**Security.** P1 thought Binance was secure for "its international reputation, advanced technology to protect users' assets, and its history of timely compensation for users' financial losses in cyber-attacks." P11 also inclined to use well-known exchanges. P14 added that attacking "big platforms" was much harder, and they were less likely to vanish. On the contrary, one of P10's friends deposited all his/her assets into an unknown exchange, and one day the exchange disappeared.

Several participants had concerns for security of CEXes. For example, P4 talked about server breakdowns in Huobi, especially when Bitcoin price went up sharply, and the trading volume exploded. Another concern raised by our participants was asset security during C2C transactions. P12 talked about the burden of telling whether the sellers were legitimate, and preferred trading with the exchanges directly, which was unavailable in CEXes.

**Privacy.** Several participants mentioned possible privacy violation in the KYC process, in which first-time users had to disclose their ID card number and mobile phone number. Several participants valued their privacy and were angry about the frequent privacy violation in China. For example, P15 complained, "I feel that I have no privacy, though they claim to protect my privacy. I'm angry, and mentally numb. I'm getting so used to it." P11 said, "Chinese people don't give a shit about privacy, but I do. Whatever platform you register, they'll check your background information. This makes me uncomfortable, but you know, I'm in China."

Often, if one needed to use an app, she had no choice but to give out her personal information. P1 acknowledged, "I actually don't like it, but I have no choice but to accept it, so that I can use these exchanges." P6 added, "Even if you don't like it, you have no choice. You want to play this game." Although P8 was worried that the government would see her transactions, and affect her social credit score (Creemers, 2018), she gave out her face ID and ID card number anyways, because she desperately wanted to buy cryptocurrency as an investment.

Interestingly, several participants thought collecting private user information could add to exchange security, as P2 said, "It'll be easier for them to trace the transactions, like this money is from whom and whom." Similarly, P9 regarded KYC as a process required in all financial business to assure asset safety, and P7 thought by providing his ID card number, he could easily reset his password.

Another scenario where privacy could be compromised was C2C transaction, where users bought cryptocurrency using fiat currency from sellers in the exchanges, often via debit card and WeChat/Alipay. Sellers often asked to see buyers' recent transaction history to check if there was invalid/illegal money, which severely violated users' privacy, as noted by P1: "Every time you trade with them, they want to see your recent transactions in your bank account. And that's privacy and I feel awkward showing them. But everyone requires them, and I have no choice."

The most apparent consequence of such privacy leakage was frequent scam phone calls, as in the case of P4 and many others.

**CEX vs. DEX.** While all our participants mainly used CEXes, we introduced DEXes to them as anonymous, regulation-free exchanges which charged more transaction fees than CEXes. We then asked them whether they would turn to DEXes.

A few participants preferred DEXes after knowing this concept. For example, P3 thought DEXes were safer than CEXes. P10 liked the convenience of DEXes: "One could use DEXes as long as she connects her wallet to them." P12 valued privacy provided by DEXes. P9 wanted more freedom over her own assets, and did not want any state regulation: "I just like the concept of decentralization. I value free rights, and that's why I choose cryptocurrency."

On the contrary, P2 regarded decentralization as "unnecessary and useless" for her own investment. P6 believed there was no way to escape regulation in China: "There's no big difference between CEXes and DEXes. All your behavior is transparent to the state." Usability was a key issue for many. For example, P11 did not choose to use DEXes because she thought the documentations were too technical to follow for novice users. P13 valued the low transaction fee and usability of CEXes, and did not view privacy as a priority, because his "personal information was already given away when registering CEXes." P10 chose CEXes because they did not have the slippage mechanism as in DEXes, and prices were consistent at the same time, thus would not cause financial losses to users.

## DISCUSSION

With an interview study, we provided a holistic view of user experience and security perception of cryptocurrency exchanges, especially CEXes, adding to the existing literature on blockchain usability (Voskobojnikov et al. 2021; Jang et al. 2021). Further, we partially explain why CEXes are preferred over DEXes by Chinese users. Generally, CEXes are valued by their good user experience, learnability, and low transaction fee, yet suffer from heavy governmental regulation (restricted internet access, prohibited C2C transaction), security vulnerabilities (server breakdown, unreliable C2C transactions), and privacy leakage during the KYC and C2C process. People's privacy perceptions toward CEXes vary, but most valued usability and low transaction fee more than privacy and decentralization.

### Country-specific Mentality of Decentralization and Privacy

Trading decentralized cryptocurrency with centralized exchanges, which are regulated, is a paradox, which intrinsically degrades the level of decentralization of cryptocurrency. For example, Chinese users can only access CEXes with a VPN, which is feasible for our participants who are relatively young and educated, but may not be possible for the majority of people who are not able to bypass the Great Firewall of China (Ensafi et al. 2015). Such a compromise of decentralization is viewed as acceptable by many of our participants.

Country-specific mentalities toward (de)centralization and privacy are an interesting phenomenon. Compared to findings in (Sas and Khairuddin, 2017), our participants similarly favor reputation, ease of use, and low transaction fee of exchanges, but put much less emphasis on decentralization and freedom from regulation. Many of the participants from China regard sensitive information leakage, such as providing phone number and ID card number during the KYC process, as less severe an issue, because they have been used to this. They claim to have no choice but to give out personal information, to use mobile applications. Another interesting opinion is that: many think they cannot escape governmental regulation even if they turn to DEXes, showing the perceived effectiveness of the regulation system. Only a few participants valued decentralization and privacy.

**Implications for Design**

Decentralization is the vision and ultimate goal of cryptocurrency, yet is hard to achieve due to the cost of decentralization (gas fee), and the high barrier for novice users to interact with the underlying blockchain. As a result, users incline to use CEXes, which are perceived more usable and cheaper than DEXes.

Similar to the finding of (Jang et al. 2021), our participants also point out the advantage of CEXes over DEXes regarding usability. CEXes are described as easy to use, rich in educational resources, resembling stock exchanges, having intuitive and familiar UI, etc. On the other hand, DEXes provide better privacy protection and are regulation-free. Thus, it is a promising direction to combine usability of CEXes and decentralization of DEXes, as insightfully put by P5. For example, DEXes can be built with CEX-like UI, which embraces users' habit in traditional financial services such as stock exchanges.

High gas fee and transaction fee are main concerns of our participants for DEXes. Some of them are not willing to "pay for decentralization" and indicate that they would only turn to DEXes after the fees are lowered. One possible way to lower transaction and gas fees is to build DEXes on Layer-2 blockchains (Stark, 2018), which are solutions for the high gas fee and scalability issue of Ethereum.

While several participants see providing their personal information such as ID card number as a way to recover their account, DEX designers could also think of effective ways to recover seed phrases or private keys (Pal et al. 2021).

## CONCLUSION

In this paper, we report results of an empirical study on user experience and security perception of cryptocurrency exchanges. While all our participants mainly use CEXes for cryptocurrency trading, we find that CEXes often provide better usability and cheaper transaction fees than DEXes. In addition, many of these Chinese users do not put an emphasis on their own privacy, and feel acceptable for their privacy to be compromised when using CEXes heavily regulated by the government. Based on these findings, we reflect on country-specific mentalities on privacy and decentralization and propose design implications toward more decentralized cryptocurrency. Future work could consider large-scale survey studies to understand cross-cultural security/privacy perceptions of cryptocurrency exchange users.